\documentclass[twocolumn,showpacs,superscriptaddress,preprintnumbers,amsmath,amssymb]{revtex4}

\usepackage{graphicx}
\usepackage{dcolumn}
\usepackage{bm}
\usepackage[perpage]{footmisc}

\begin{document}



\title{Performance of the veto detector incorporated into the ZEPLIN--III experiment}%

\author{C.~Ghag\footnote{corresponding author: c.ghag@ed.ac.uk}} \affiliation{School of Physics \& Astronomy, University of Edinburgh, UK}
\author{D.Yu.~Akimov}\affiliation{Institute for Theoretical and Experimental Physics, Moscow, Russia}
\author{H.M.~Ara\'{u}jo}\affiliation{Blackett Laboratory, Imperial College London, UK}
\author{E.J.~Barnes}\affiliation{School of Physics \& Astronomy, University of Edinburgh, UK}
\author{V.A.~Belov} \affiliation{Institute for Theoretical and Experimental Physics, Moscow, Russia}
\author{A.A.~Burenkov} \affiliation{Institute for Theoretical and Experimental Physics, Moscow, Russia}
\author{V.~Chepel}\affiliation{LIP--Coimbra \& Department of Physics of the University of Coimbra, Portugal}
\author{A.~Currie}\affiliation{Blackett Laboratory, Imperial College London, UK}
\author{L.~DeViveiros}\affiliation{LIP--Coimbra \& Department of Physics of the University of Coimbra, Portugal}
\author{B.~Edwards}\affiliation{Particle Physics Department, STFC Rutherford Appleton Laboratory, Chilton, UK}
\author{V.~Francis}\affiliation{Particle Physics Department, STFC Rutherford Appleton Laboratory, Chilton, UK}
\author{A.~Hollingsworth} \affiliation{School of Physics \& Astronomy, University of Edinburgh, UK}
\author{M.~Horn}  \affiliation{Blackett Laboratory, Imperial College London, UK}
\author{G.E.~Kalmus} \affiliation{Particle Physics Department, STFC Rutherford Appleton Laboratory, Chilton, UK}
\author{A.S.~Kobyakin} \affiliation{Institute for Theoretical and Experimental Physics, Moscow, Russia}
\author{A.G.~Kovalenko} \affiliation{Institute for Theoretical and Experimental Physics, Moscow, Russia}
\author{V.N.~Lebedenko}  \affiliation{Blackett Laboratory, Imperial College London, UK}
\author{A.~Lindote} \affiliation{LIP--Coimbra \& Department of Physics of the University of Coimbra, Portugal}
\affiliation{Particle Physics Department, STFC Rutherford Appleton Laboratory, Chilton, UK}
\author{M.I.~Lopes} \affiliation{LIP--Coimbra \& Department of Physics of the University of Coimbra, Portugal}
\author{R.~L\"{u}scher} \affiliation{Particle Physics Department, STFC Rutherford Appleton Laboratory, Chilton, UK}
\author{K.~Lyons}  \affiliation{Blackett Laboratory, Imperial College London, UK}
\author{P.~Majewski} \affiliation{Particle Physics Department, STFC Rutherford Appleton Laboratory, Chilton, UK}
\author{A.St\,J.~Murphy} \affiliation{School of Physics \& Astronomy, University of Edinburgh, UK}
\author{F.~Neves} \affiliation{LIP--Coimbra \& Department of Physics of the University of Coimbra, Portugal}
\affiliation{Blackett Laboratory, Imperial College London, UK}
\author{S.M.~Paling}\affiliation{Particle Physics Department, STFC Rutherford Appleton Laboratory, Chilton, UK}
\author{J.~Pinto da Cunha} \affiliation{LIP--Coimbra \& Department of Physics of the University of Coimbra, Portugal}
\author{R.~Preece} \affiliation{Particle Physics Department, STFC Rutherford Appleton Laboratory, Chilton, UK}
\author{J.J.~Quenby}  \affiliation{Blackett Laboratory, Imperial College London, UK}
\author{L.~Reichhart} \affiliation{School of Physics \& Astronomy, University of Edinburgh, UK}
\author{P.R.~Scovell} \affiliation{School of Physics \& Astronomy, University of Edinburgh, UK}
\author{C.~Silva} \affiliation{LIP--Coimbra \& Department of Physics of the University of Coimbra, Portugal}
\author{V.N.~Solovov} \affiliation{LIP--Coimbra \& Department of Physics of the University of Coimbra, Portugal}
\author{N.J.T.~Smith} \affiliation{Particle Physics Department, STFC Rutherford Appleton Laboratory, Chilton, UK}
\author{P.F.~Smith} \affiliation{Particle Physics Department, STFC Rutherford Appleton Laboratory, Chilton, UK}
\author{V.N.~Stekhanov} \affiliation{Institute for Theoretical and Experimental Physics, Moscow, Russia}
\author{T.J.~Sumner} \affiliation{Blackett Laboratory, Imperial College London, UK}
\author{C.~Thorne} \affiliation{Blackett Laboratory, Imperial College London, UK}
\author{R.J.~Walker} \affiliation{Blackett Laboratory, Imperial College London, UK}

\date{\today}

\begin{abstract}
\noindent The ZEPLIN--III experiment is operating in its second phase at the Boulby Underground Laboratory in search of dark matter WIMPs.  
The major upgrades to the instrument over its first science run include lower background photomultiplier tubes and installation of a plastic scintillator veto system.  
Performance results from the veto detector using calibration and science data in its first six months of operation in coincidence with ZEPLIN--III are presented.  
With fully automated operation and calibration, the veto system has maintained high stability and achieves near unity live time relative to ZEPLIN--III.  
Calibrations with a neutron source demonstrate a rejection of 60\% of neutron-induced nuclear recoils in ZEPLIN--III that might otherwise be misidentified as WIMPs.  
This tagging efficiency reduces the expected untagged nuclear recoil background from neutrons during science data taking to a very low rate of $\simeq$0.2 events per year in the WIMP acceptance region.  
Additionally, the veto detector provides rejection of 28\% of $\gamma$-ray induced background events, allowing the sampling of the dominant source of background in ZEPLIN--III -- 
multiple scatter $\gamma$-rays with rare topologies.  
Since WIMPs will not be tagged by the veto detector, and tags due to $\gamma$-rays and neutrons are separable, 
this population of multiple scatter events may be characterised without biasing the analysis of candidate WIMP signals in the data.

\end{abstract}


\maketitle

\section{INTRODUCTION}

\noindent The ZEPLIN--III instrument~\cite{akimov2007,z3sim} is a two phase (liquid/gas) xenon detector with photomultiplier (PMT) readout designed to observe the low energy nuclear recoils from elastic scattering 
of weakly interacting massive particles (WIMPs).  ZEPLIN--III records both direct scintillation light ({\em S1}) and electroluminescence from ionisation ({\em S2}) of the xenon target~\cite{lebedenko} 
following an energy deposition.  The ratio of the signal strength in these channels differs for electron and nuclear recoil interactions, 
allowing discrimination between incident particle species and the efficient rejection of most background events.  
However, as is the case for all direct dark matter search experiments, single elastic scattering of neutrons and WIMPs generate the same signature, rendering the former a particularly problematic background.  
Furthermore, even where electron recoil discrimination is excellent, it is probabilistic and there always remains a small probability for those electron recoils to be misidentified as nuclear recoils.  
An external plastic scintillator based veto detector can help decrease these backgrounds by removal of coincident $\gamma$-ray and neutron induced nuclear recoil events (WIMPs are extremely unlikely to scatter in both detectors).  
In addition, by providing an independent measurement of $\gamma$-ray and neutron rates in the WIMP target, 
it will also help decrease the systematic uncertainty in the estimation of these background rates, 
thereby increasing the significance of a non-zero observation.  
Experiencing the same physical conditions as ZEPLIN--III, the veto system also provides valuable environmental diagnostic information as well as independent {\em in situ} measurements 
of ambient background radioactivity levels.

The first science run of ZEPLIN--III resulted in the exclusion, with 90\% confidence, of spin-independent WIMP-nucleon cross sections above 8.1$\times$10$^{-8}$~pb
for a WIMP mass of 55 GeV/c$^2$~\cite{z3fsr}, and spin-dependent WIMP-neutron cross sections above 1.8$\times$10$^{-2}$~pb~\cite{z3fsrsd}.  
The explanation of the DAMA annual modulation signal~\cite{DAMA} by inelastic dark matter~\cite{iDM} in a 
Maxwellian halo scattering on iodine was ruled out with 87\% confidence~\cite{z3inelastic}.  
Since then the instrument has undergone an upgrade with the replacement of the internal PMT array, the dominant source of background during the first science exposure.  
The addition of the veto detector will further improve the sensitivity of a second run of the experiment.  
These upgrades are as originally conceived, planned and scheduled prior to the first science run.  
As such, the necessity for a veto detector with high neutron rejection efficiency has been motivated by conservative nuclear recoil expectation levels from the first science run.  
Since considerably reduced second science run neutron induced nuclear recoil predictions have been successfully met following the internal PMT array replacement, 
the diagnostic capability and $\gamma$-ray tagging efficiency of the veto detector become increasingly significant.

By design, the veto detector itself contributes negligible additional background to the xenon target.  
All detector components have had their performance characterised extensively to maximise the expected detection efficiencies for both electron and nuclear recoils.  
These studies, supplemented with Monte Carlo simulations using the GEANT4 toolkit~\cite{geant4}, have been presented previously~\cite{vetoPaper1}.  
In the following sections we present the realised performance of the veto detector following six months of continuous operation.  
We first give a brief description of the system followed by results from calibrations showing detector stability and operational robustness.  
In the subsequent sections the efficiencies for rejecting coincident $\gamma$-ray and neutron events in ZEPLIN--III are presented.  
These so-called `tagging' efficiencies are defined as the fractions of events occurring in the xenon target that are successfully identified and may be vetoed.  
As is described in Section~\ref{gammatagging}, $\gamma$-rays that deposit energy in the liquid xenon and the veto detector do so approximately simultaneously.  
Consequently, such coincident vetoed events are labelled `prompt' tags.  
Signals in the veto following neutron recoils in ZEPLIN--III, however, arrive after a time which depends on the passive shielding and veto detector geometry and composition.  
The majority of these neutron events are well separated in time from prompt tags and, as is described in Section~\ref{neutrontagging}, are labelled `delayed' tags.  
In Section~\ref{implications} we discuss briefly how the tagging of a unique background adds to the statistical evidence of a signal appearing among the untagged events.

\section{Detector Description}

\begin{figure}[ht]
\includegraphics[width=8.6cm]{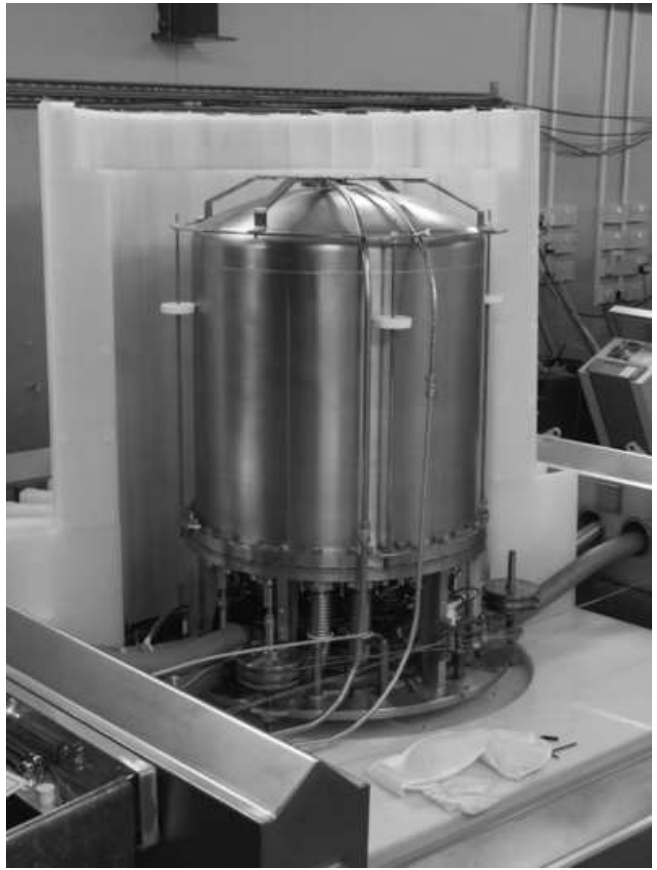}
\caption{\label{vetofigure1}An image of the passive Gd-loaded polypropylene sections of the veto detector partially assembled around the ZEPLIN--III instrument.  
Thirty-two such sections form a closed barrel and a roof plug is lowered onto the barrel to enclose ZEPLIN--III completely. 
The active scintillator modules are then placed around the polypropylene, and make up in total 30~cm thickness hydrocarbon shielding.}
\end{figure}

\begin{figure}[h]
\includegraphics[width=8.6cm]{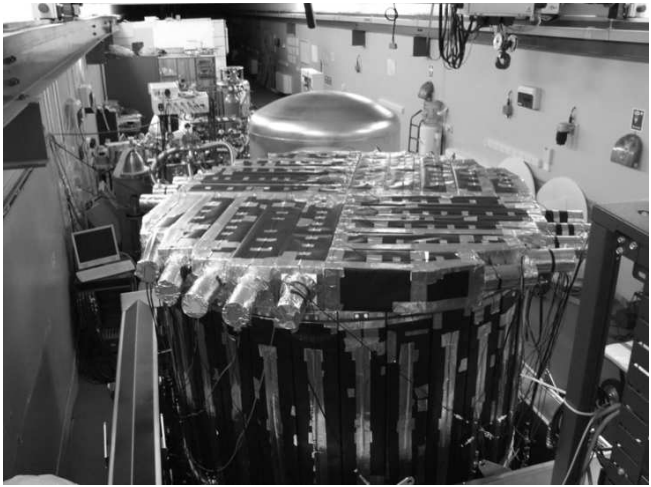}
\caption{\label{vetofigure2}An aerial view of the completed veto detector before the Pb castle is built up around the entire assembly.  
The Gd-loaded polypropylene sections have been completely enclosed by the scintillator modules.  Thirty-two sections make up the barrel with a further twenty modules 
on the roof above ZEPLIN--III.  The enclosures for the PMTs attached to the scintillators on the roof can be seen whereas those on the standing barrel scintillators 
are in recesses built into the polypropylene sections.  The outer diameter of the barrel is 160~cm.}
\end{figure}

\begin{figure}[h]
\includegraphics[width=8.6cm]{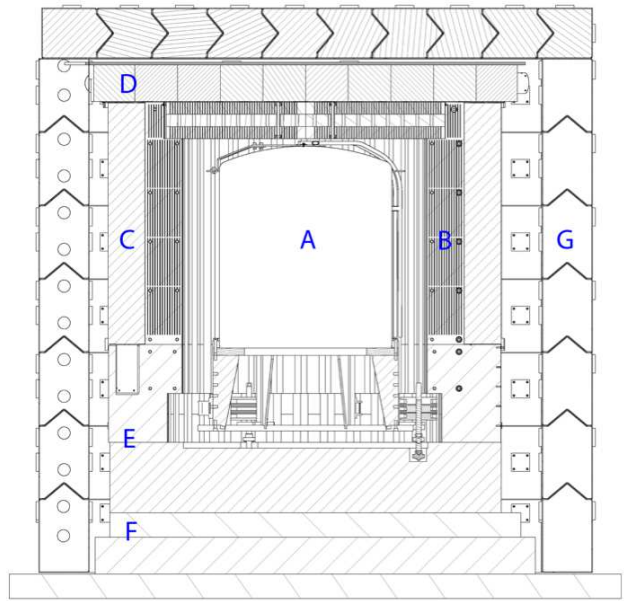}
\caption{\label{cad}Cross-sectional view of the ZEPLIN-III experiment in its second science run configuration.  
In the centre is the ZEPLIN-III detector (labelled A) showing the outer copper vacuum vessel.  
Forming a barrel around ZEPLIN-III are the 32 Gd-loaded polypropylene pieces and roof plug (labelled B and hatched).  
Surrounding these are the active scintillator modules (C) with PMTs housed in cups and recessed into the lower polypropylene structure.  
The 20 roof modules of scintillator (D) rest on the roof plug.  
The lower polypropylene structure (E) contains no Gd and rests on a copper and a lead base (F).  
Finally, the lead castle (G) envelopes the entire assembly.}
\end{figure}

The veto instrument has two main components.  
A Gd-loaded polypropylene structure, close-fitting around and above ZEPLIN--III, moderates and captures neutrons that may have scattered in the WIMP target; 
this structure is shown in Figure~\ref{vetofigure1} during assembly.  
The capture $\gamma$-rays are detected by 52 plastic scintillator modules (polystyrene-based UPS–-923A) which are assembled around the polypropylene; 
these are shown in Figure~\ref{vetofigure2} after assembly.  
The entire structure is subequently enclosed in a Pb castle, as depicted in Figure~\ref{cad}, a schematic drawing of ZEPLIN--III, the veto system, and passive shielding.  
A detailed description of the veto detector hardware is given in Ref.~\cite{vetoPaper1}, here only a brief summary is presented.  

The scintillator modules each have one PMT (ETEL-9302KB) optically coupled to one end.  
The PMT and associated electronics are housed in a plastic cylinder which is bonded chemically to the scintillator; some such tubes are visible on the roof section in Figure~\ref{vetofigure2}.  
To minimise loss of light within a scintillator bar, each piece has been wrapped along its length with PTFE sheet of high diffuse reflectivity, and then in black opaque sheet to provide light tightness.  
A reflector (aluminised Mylar film) is placed at the far end from the PMT to increase the light collection of the unit.  
All scintillator modules are 15~cm thick.  
Modules on the roof vary in length from $\sim$40~cm to $\sim$100~cm to form an approximate disc whereas the trapezoidal shaped barrel modules are all 100~cm in length.

The scintillator modules of the barrel surround individual 15~cm thick polypropylene sections that are loaded with 0.4\% Gd by weight.  
The active and passive plastic sections together maintain the 30~cm hydrocarbon shielding thickness of the first science run.  
A single Gd-loaded polypropylene roof plug supports the 20 roof modules of scintillator.  
At 0.4\% concentration of loading $>$99\% of neutrons moderated to thermal energies undergo radiative capture on $^{157}$Gd (natural abundance 15.7\%) 
which has an extraordinarily high capture cross section of 2.4$\times$10$^{5}$~barn~\cite{gdcrosssec}.  
The benefit of using Gd is two-fold.  
Firstly, the states populated in $^{158}$Gd by the neutron capture decay with the emission of up to 8~MeV distributed amongst typically 3--4 $\gamma$-rays 
rather than the single 2.2~MeV $\gamma$-ray that follows radiative capture on H.  
The inclusion of Gd, thus,  increases significantly the neutron tagging efficiency.  
Secondly, the mean capture time is reduced relative to that on H.  This allows adoption of a shorter coincidence window 
resulting in reduced data volume and a lower accidental coincidence rate.  
The Gd, in the form of Gd$_{2}$O$_{3}$, is mixed into an epoxy and set into slots in the polypropylene sections, with slot pitch and width 10~mm and 2~mm, respectively.  
The veto detector and the ZEPLIN--III instrument stand on a 30~cm thick polypropylene base and are enclosed within a 20~cm Pb castle.  
A single blue LED transmits light through 52 optical fibres to the far end of each of the veto detector modules for weekly monitoring of the scintillator/PMT response.  
Each PMT output is digitised with 14-bit resolution at 100~MS/s for a waveform duration of 320~$\mu$s.  

The operation of the veto detector allows for a combination of `slave' and `master' modes.  
In {\em slave} mode, timelines from all modules are recorded when an external trigger from ZEPLIN--III is received.  
In {\em master} mode, when certain conditions are met, the veto system triggers independently.  
In the present configuration this happens when three or more modules each register at least 10 photoelectrons in the same event.  
This threshold results in an event rate of approximately 2~Hz 
and allows for an independent measure of background from the detector surroundings, especially ambient neutrons from the laboratory rock.

Finally, a muon trigger is derived from the roof modules with no condition on multiplicity ({\em i.e.}, the number of modules with simulataneously occurring pulses). 
This requires an external triggering unit in order to avoid the large data volumes otherwise generated by a software trigger testing individual modules.  
In addition to providing a direct measure of the atmospheric muon flux through the laboratory, the purpose of the unit is to help determine the 
muon-induced neutron production rate in Pb, a measurement of interest to the rare-event search community.  
Using the Pb castle as a target and taking advantage of the low energy threshold of the veto detector 
as well as its segmented design to provide angular information, it is hoped that an improvement upon previous measurements conducted in a similar fashion may be made~\cite{muonal}.
Results of these studies will be presented at a later date.  
Here we focus on the results from the system in its primary role of a veto detector.

\section{Detector Stability}

Daily calibration of the veto detector with a radioactive $\gamma$-ray source is impractical: besides taking too long to calibrate all 52 modules, 
a high $\gamma$-ray flux inside the castle might compromise the stability of ZEPLIN--III.  
To probe the stability of the veto system we monitor, separately, the light transmission of the scintillators (using the fibre-coupled LED signal) and the electronic gain (by measuring the single photoelectron (SPE) response of the PMTs).  
In addition, we confirm the rate of coincidences with ZEPLIN--III on a daily basis, record the rate and energy spectrum of the master-mode triggers, and measure the $\gamma$-ray tagging efficiency and a number of environmental parameters which can be correlated with the data {\em a posteriori}.  
The PMT and LED calibrations are described next, followed by the methods used to synchronise the two instruments, looking in particular at how these evolved over several months of underground operation.

\subsection{\label{spemeasurements}SPE calibration}

An automated daily SPE analysis examines the preceding day's dataset (excluding the portion of the waveforms where a signal in prompt coincidence with ZEPLIN--III would appear) 
and searches for the smallest area pulses found above the baseline.  
This is done by the pulse-finding software `raVen', a dedicated code derived from the ZEPLIN--III data reduction application ZE3RA~\cite{ze3ra}.  
Most of the 320~$\mu$s long timelines are empty save for signals from spontaneous photoelectron emission from the PMTs, due to both dark noise and low-level light leakage.  
The peak position of the resulting SPE spectrum from each channel, which is very well defined for all modules, 
has been used to equalise the PMT gains across the array prior to the science run.  

\begin{figure}[ht]
\includegraphics[width=8.0cm]{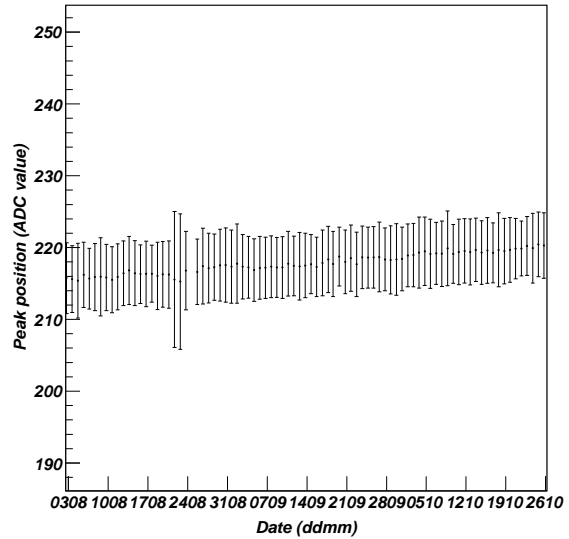}
\caption{\label{speTrend}Evolution of the mean of the SPE peak positions for all 52 veto detector PMTs over a 3 month period.  
The bars are the r.m.s. of the 52 means of the SPE peak positions.}
\end{figure}

The SPE analysis ensures stability of the gains as well as providing a conversion from pulse size to number of photoelectrons, allowing accurate calibration of the signal response.  
Calibrating using SPEs from the science acquisition itself, rather than a dedicated SPE dataset, avoids interrupting the science run and maximises the WIMP search exposure.  
Moreover, this ensures signal pulses are calibrated with contemporaneous SPE data subjected to identical operational conditions.  
Figure~\ref{speTrend} shows the average of the mean SPE values from all 52 PMTs.  
With continuous dark operation of the PMTs, their gains are found to be rising on average by $\sim$0.6\% per month in the first 3 months of operation.  
With the PMTs on the outside of the plastic shielding, it is expected that they should be affected by seasonal environmental changes in the underground laboratory.  
Although this effect is small, it nonetheless highlights the need for this regular calibration.  

\subsection{LED calibration}

\begin{figure}[ht]
\includegraphics[width=8.0cm]{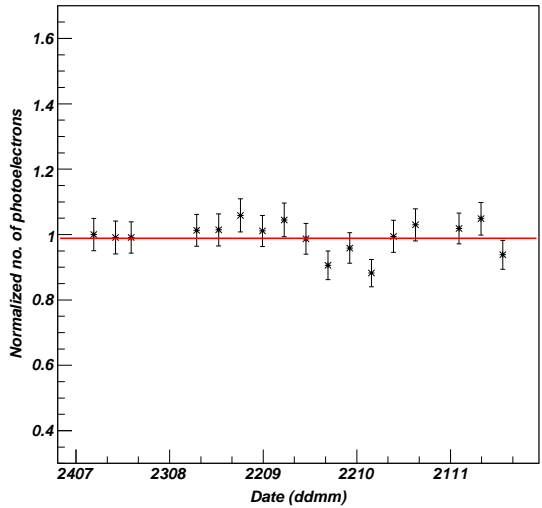}
\caption{\label{ledTrend}Evolution of the number of photoelectrons from LED exposures (normalised to the first measurement) for a single typical veto module.}
\end{figure}

The response of the PMTs to light transmission through the plastic scintillator is measured with weekly illumination from a blue LED 
coupled through an independent fibre optic cable to each module.  
Datasets are acquired during a brief pause in the running of ZEPLIN--III used to perform maintenance operations such as filling of the LN$_{2}$ reservoir.  
The LED is mounted in an acrylic light guide from which 52 optical fibres emerge.  These fibres are secured in recesses in the plastic scintillators at the opposite ends to the PMTs.  
The LED is pulsed at a frequency of approximately 30~Hz for 300~s and light emitted from the fibres is transmitted through the scintillator to the PMT photocathode, 
generating for each pulse an average of 48 photoelectrons in each module.  
A relatively low rate is used so as not to distort spectral shape through induction of afterpulsing or saturation effects in the PMTs.  
The number of photoelectrons contributing to the peak LED response is calculated from the SPE peak position and estimated from the width of the distribution for consistency.  
The normalised number of photoelectrons detected in each module is then tracked as a function of time throughout the course of the experiment 
and is plotted in Figure~\ref{ledTrend}, showing no detectable change in optical transmission so far.  
A full study of the properties of UPS--923A scintillator, including results from dedicated measurements of its nuclear quenching factor for low energy recoils, will be presented elsewhere~\cite{leaQF}.  

\subsection{ZEPLIN--III - Veto event synchronisation}

The ZEPLIN--III and veto detector data acquisition systems operate at different sampling speeds and digitise waveforms of different duration with different resolutions.  
Accurate target-veto timing is achieved using a bespoke synchronisation unit clocked at 1~MHz.  
This sends a 32-bit digital stamp to both acquisition systems following a trigger from either detector.  
A number of additional methods have been implemented for redundancy.  
This is critical since, in case of malfunction of the synchronisation unit, the veto efficiency might be compromised for a significant part of the dataset.  
Firstly, an internet time server is used to synchronise the two data streams to 4~ms.  
It is then required that the time difference between consecutive coincident events ({\em i.e.}, excluding veto self triggers) agrees to within 1~ms; 
since the target trigger rate is only $\sim$0.4~s$^{-1}$ and all genuine coincidences have triggers sourced from ZEPLIN--III, the event time distribution proves to be a powerful selector.  
Finally, the summed signal of the PMT response of ZEPLIN--III is digitised into the veto data acquisition and pulse parameters must agree.  
Using these methods the fraction of unsynchronised events is negligible.

\section{Detector Interaction Rates}\label{signalrates}

\begin{figure}[ht]
\includegraphics[width=8.6cm]{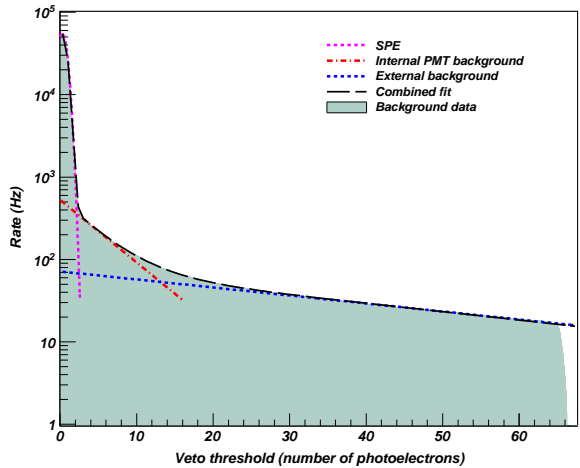}
\caption{\label{veto_rate_full}Cumulative background signal rate in the full veto detector array as a function of photoelectron threshold.  
All pulses are considered irrespective of multiplicity.  
Statistical errors are too small to be seen on this scale.  
The pulses are measured from slave mode triggerd events and the rate is truncated at 65 photoelectrons where pulse height saturation takes effect.  
The fit to the data is composed of three well resolved components made up of SPEs, internal background from the PMTs and $\gamma$-ray background from 
radiological contamination within surrounding materials and the veto detector itself.}
\end{figure}

The integral signal rate in the veto detector during science data-taking is shown in Figure~\ref{veto_rate_full}.  
This is the cumulative distribution of pulse sizes, given in total number of photoelectrons, as a function of threshold.  
Note that there may be several pulses per event, the latter being determined by a single data acquisition trigger.  
As such, Figure~\ref{veto_rate_full} indicates the actual background pulse rate as measured by the veto detector.  
The plot extends to the point of pulse saturation, which is due to the full range of the digitiser ADCs and the gain settings of the PMTs, 
and occurs at approximately 65 photoelectrons (equivalent to several MeV of energy deposition across the array).  
The pulse rate from background data is well described by a combined fit of three components.  
The first is a semi-Gaussian fit to the SPE peak up to 2 photoelectrons, 
characterising thermionic emission from the photocathodes and ambient light leakage into the scintillators.  
As described in Section~\ref{gammatagging}, such single pulses are considered below threshold for valid coincident events.  
The second component is an exponential fit to background from radiological contamination from within the veto detector PMTs.  
In particular, the $^{40}$K content dominates between 2-15 photoelectrons.  
$^{40}$K has an 89\% $\beta^{-}$ decay branching ratio.  
For a refractive index of 1.49 at 400~nm, $\beta^{-}$ radiation emitted from the potassium generators in the PMT behind the photocathode will produce Cherenkov photons 
in the window when the electron energy exceeds 178~keV.  
Dominated by signal from these photons and supplemented by Bremsstrahlung radiation in the window and the scintillator from $\beta^{-}$ particles with energies in the tail of the distribution that escape the window, 
the slope and magnitude of the contribution from this component are consistent with expectations~\cite{wright}.  
The third component in the combined fit to the pulse rate spectrum dominates at higher energies and is of particular relevance to ZEPLIN--III.  
It is an exponential fit to the background component due to $\gamma$-ray radioactivity from U and Th decay chains and $^{40}$K $\gamma$-rays within shielding and surrounding materials interacting within the scintillator.  
This includes the contributions from all of the veto detector components such as the plastic and the PMTs, 
and is dominated by both plastics, despite their low radiological content, due to their high mass.  
The measured $\gamma$-ray background agrees with predictions from Monte Carlo simulations presented in Ref.~\cite{vetoPaper1} 
provided that one adopts contamination levels just consistent with the sensitivity of the radio-assays that produced null results, most notably the Gd-loaded polypropylene.  
This is also consistent with the $\gamma$-ray background observed in the liquid xenon target~\cite{backgrounds}.  
The radiological activity from the veto detector will contribute less than 0.01 single neutron elastic scatters per year in the ZEPLIN--III fiducial volume.  
$\gamma$-ray emission from radioactive impurities in the plastic scintillator also poses a low risk since it is tagged with near unity efficiency.

The background signal rate in the veto detector is also of direct consequence to its performance since this defines the probability of an event in ZEPLIN--III being accidentally vetoed.  
The accidentals rate is determined separately for prompt $\gamma$-ray tagging and for delayed neutron tagging since the selection criteria 
(such as energy threshold, pulse arrival time, multiplicity, etc) for a coincident event to be labelled as a $\gamma$-ray or neutron differ.  

\begin{figure}[ht]
\includegraphics[width=8.6cm]{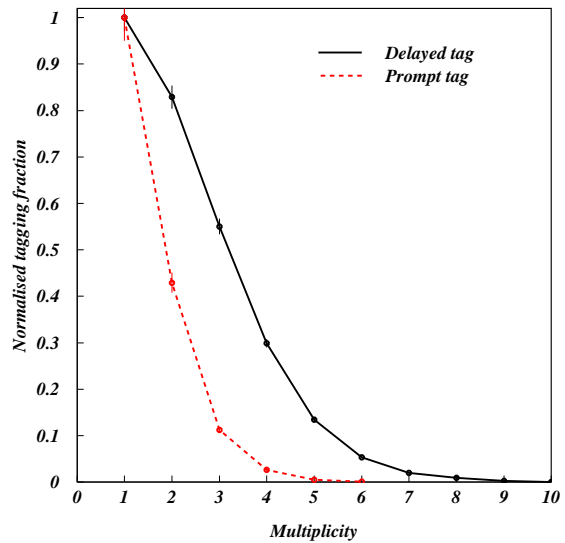}
\caption{\label{nvg_norm}The relative efficiencies of the prompt $\gamma$-ray and delayed neutron tags, as a function of module multiplicity.  
The thresholds for selection of prompt and delayed events are as described in Sections~\ref{gammatagging} and~\ref{neutrontagging}.  
It is solely the multiplicity requirement that is varied.  
The maximal efficiencies have been normalised to unity to provide a direct comparison.  
Lines through the points have been added to guide the eye.  
The delayed neutron tag varies considerably more slowly with increasing multiplicity than the prompt tag.  
This results from the former being dominated by detection of multiple high energy $\gamma$-rays following de-excitation of the $^{158}$Gd nucleus.}
\end{figure}

In principle, the signal multiplicity could be an important characteristic with which to discriminate between veto events caused by $\gamma$-ray and by neutron backgrounds, 
due to the multiple, several MeV $\gamma$-rays that result from radiative capture on the Gd loading.  
Figure~\ref{nvg_norm} shows the multiplicity distributions of tagged $\gamma$-ray background signals in the veto detector and of tagged neutron events from calibration data.  
That the tagging efficiency for background $\gamma$-rays falls more rapidly with multiplicity than for delayed neutron signals is clearly illustrated.  
However, as is explored in Section~\ref{neutrontaggingefficiency}, exploitation of this difference was not found to be beneficial for the present application, 
although it might be a useful characteristic in other applications.  
Successful algorithms for $\gamma$-ray and neutron tagging have been found that use only timing and energy deposition information; these are now described. 

\section{\label{gammatagging}$\gamma$-ray tagging}

Following a $\gamma$-ray Compton scatter in the xenon target, the energy deposition from an electron recoil will trigger both ZEPLIN--III and veto detector data acquisitions.  
The scattered $\gamma$-ray may interact promptly within the veto detector and only a narrow coincidence window is required to identify these $\gamma$-rays.  
The criteria for designation of such `prompt tags' are described below.  
These include determining the optimal prompt coincidence window and setting of an appropriate veto detector threshold (in energy and multiplicity) 
to maximise acceptance for genuine prompt events whilst minimising those that are accidentally coincident.  
This is followed by the resulting tagging efficiency for prompt $\gamma$-ray signals.

\subsection{Selection criteria for prompt signals}

\begin{figure}[ht]
\includegraphics[width=8.6cm]{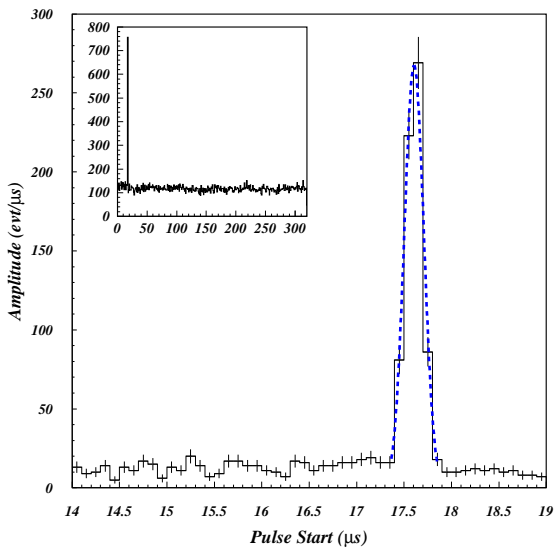}
\caption{\label{starttime_merge}Pulse time distribution from 14--19~$\mu$s in the veto detector when triggered by ZEPLIN--III.  
The peak at 17.7~$\mu$s is from pulses in prompt coincidence with the {\em S1}.  
Smaller signals in the WIMP target, which trigger on {\em S2} instead, are contained within a 16~$\mu$s region before the peak.  
Pulses occurring after this peak cannot be in prompt coincidence, and represent the background rate in the veto detector itself.
Insert: the pulse start times in the veto detector for the full 320~$\mu$s timelines recorded.}
\end{figure}

ZEPLIN--III discriminates between incident particle species by recording a prompt scintillation coming directly from the interaction site, {\em S1}, 
and a delayed electroluminescence signal caused by the ionisation escaping into the gas volume, {\em S2}.  
The delay between {\em S1} and {\em S2} can be up to $\sim$16~$\mu$s depending on the depth in the liquid at which the interaction took place.  
ZEPLIN--III derives its own trigger either from {\em S1} or {\em S2} depending on the energy deposit and hence digitises timelines $\pm$16~$\mu$s either side of the trigger.  
Consequently, the veto data acquisition must do likewise (in fact a slightly wider pre-trigger timeline is adopted, -20~$\mu$s).  
The trigger point in ZEPLIN--III corresponds to the 17.7~$\mu$s mark in the veto detector timelines.  
For {\em S1} triggered events, any prompt coincidences in the veto detector will appear as a peak at approximately 17.7~$\mu$s.  
This is illustrated in Figure~\ref{starttime_merge} showing the times of veto detector pulses recorded in background data, having been triggered in slave mode by ZEPLIN--III.  
A single peak with a FWHM of 226~ns corresponding to prompt coincidences with {\em S1} triggers is visible above background (insert) and the region around this peak has been enlarged.  
The events triggered by {\em S2} have {\em S1} signals preceding the trigger point and {\em S1} coincident pulses in the veto detector are distributed in the region prior to the prompt peak.  
The lower level of signals occurring after the prompt peak, unrelated to the trigger signals in ZEPLIN--III, are random background seen in the veto detector itself.  

\begin{figure}[ht]
\includegraphics[width=8.6cm]{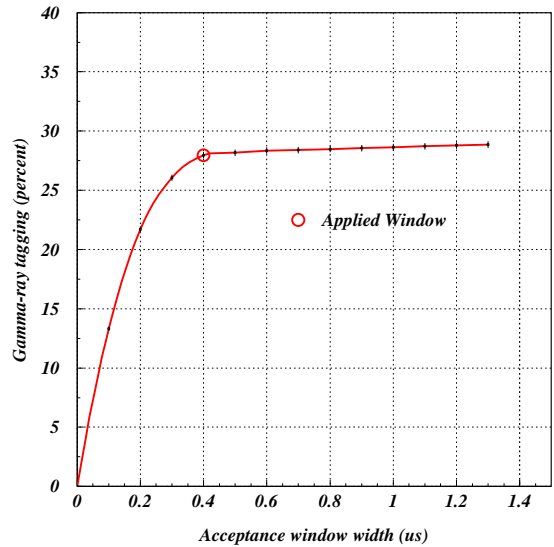}
\caption{\label{promptWindow}The efficiency for tagging prompt coincidence events between ZEPLIN--III and the veto detector, 
as a function of acceptance window width.  The efficiency rises rapidly as the window is opened to 0.4~$\mu$s, 
but beyond this any increases in efficiency corresponds to the inclusion of accidental coincidence events.}
\end{figure}

A number of factors affect the timing resolution of the target-veto combination.  
Naturally, the limited sampling speed of the veto digitisers (100~ns sampling) dominates this, as illustrated in Figure~\ref{starttime_merge}.  
In addition, there is a smaller contribution from the timing jitter in ZEPLIN--III itself (2~ns sampling), both in determining the arrival time of {\em S1} and, more importantly, 
in measuring the {\em S1}--{\em S2} separation.  
The latter applies to {\em S2} triggers and is a few tens of ns, reflecting the slower build up of the electroluminescence response in the target and charge diffusion whilst drifting to the liquid surface.  
Taking these factors into account, a coincidence window of 0.4~$\mu$s has been defined as the prompt window.  
Extending the window beyond this only increases the number of events by introducing accidental coincidences, 
which lower the effective exposure of the xenon target by unduly rejecting potential WIMP events. 
Figure~\ref{promptWindow} shows the efficiency of the veto detector for tagging prompt events rising rapidly as the coincidence window is opened to 0.4~$\mu$s. 

The rate of accidental coincidences can be calculated from the product of the event rates in both systems, and can be measured directly by applying the same window off-coincidence.  
This rate is 0.4\% for a 0.4~$\mu$s window centred on the peak in Figure~\ref{starttime_merge} using a 2 photoelectron threshold.  
Figure~\ref{promptWindow} corroborates independently the accidental coincidence rate.  
Since the plot shows the prompt $\gamma$-ray tagging efficiency values with the thresholds as described above, 
extrapolating a fit to the linear region at high acceptance window values to the y-axis results in an estimate of the tagging efficiency in the absence of accidental coincidences.  
The difference between this value and the measured tagging efficiency for any given acceptance window is then the accidental coincidence rate for that window.  
At 0.4~$\mu$s window width the accidental rate is predicted to be less than 0.4\%, in agreement with measurements.

\begin{figure}[ht]
\includegraphics[width=8.6cm]{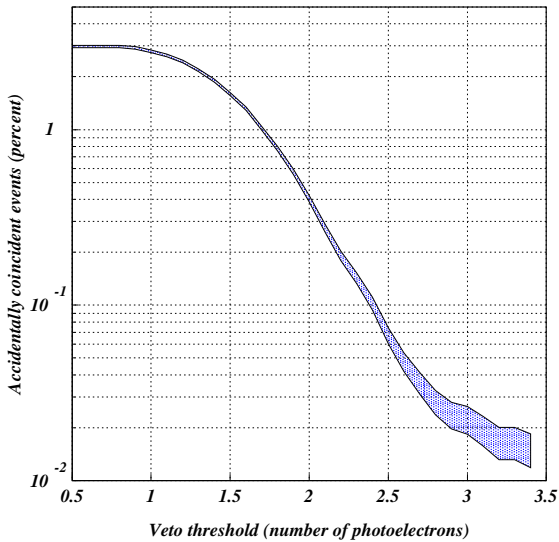}
\caption{\label{gammaaccidental}Accidentally tagged events in the 0.4~$\mu$s prompt window as a function of veto detector threshold.  
The band represents 1$\sigma$ errors.  
For the prompt tag threshold of 2 phototelectrons equivalent signal in the veto detector, 
the probability that a tagged event is not correlated with the signal in ZEPLIN--III is 0.4\%.}
\end{figure}

The prompt tag threshold requires a minimum of 2 photoelectrons equivalent pulse signal distributed in any pattern across the veto detector, 
provided that any coincident pulses across multiple modules are themselves within $\pm$0.2~$\mu$s of one another.  
Lowering the threshold any further would result in a prohibitive accidental rate due to PMT dark noise and light leakage at the single photoelectron level.  
This is illustrated in Figure~\ref{gammaaccidental}, showing the fraction of events accidentally tagged as $\gamma$-rays in ZEPLIN--III, as a function of signal threshold in the veto detector.  
Here, the veto detector channel timelines are searched for accidental coincident pulses outside of the prompt coincidence window and the pre-trigger region.  
This ensures genuine coincidences with {\em S1} signals are ignored.
 
\subsection{$\gamma$-ray tagging efficiency}

\begin{figure}[ht]
\includegraphics[width=8.6cm]{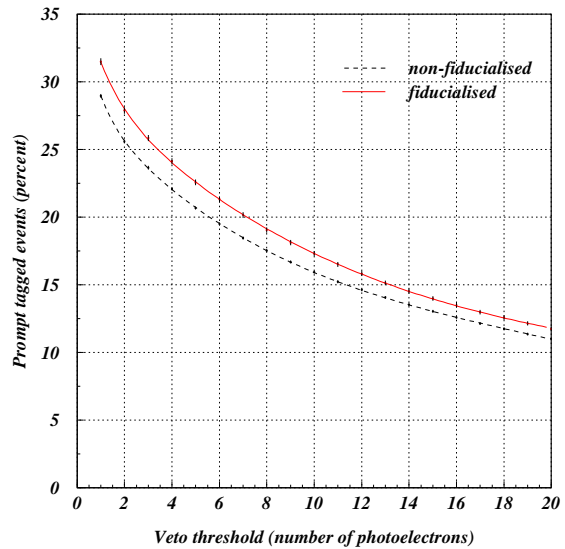}
\caption{\label{gamma_sig_comp_cg}The percentage of coincident events satisfying the prompt tag as a function of veto detector threshold, 
for all synchronised ZEPLIN--III events depositing up to 100~keV$_{ee}$ in the target xenon as well as for the fiducialised target.  
No additional cuts or restrictions have been applied to the data.}
\end{figure}

With the prompt tag criteria set out above, 
the tagging efficiency can be determined from synchronised background data triggered by ZEPLIN--III with the veto detector running in slave mode.  
The resulting prompt tagging efficiency is shown as a function of veto detector threshold in Figure~\ref{gamma_sig_comp_cg}.  
The plot shows the measured efficiencies for all triggered events below 100~keV$_{ee}$ energy in ZEPLIN--III 
(where keV$_{ee}$ is electron-equivalent energy using 122 keV $\gamma$-rays from a $^{57}$Co source to set the energy calibration).  
Also shown is the tagging efficiency for those events that occur in the fiducialised xenon of the ZEPLIN--III instrument.  
As expected, the efficiency increases when considering the central fiducial target xenon.  
Here background events are more likely to be from Compton scatters 
(rather than, for example, $\beta$-induced background or $\alpha$-particles that cannot directly give signal in the veto modules) 
and the $\gamma$-rays have consequently a higher probability of being detected in the veto modules.  
The tagging efficiency for prompt events is (28.1$\pm$0.2)\% for synchronised electron recoil background in the fiducial liquid xenon volume.  
Monte Carlo simulations of the $\gamma$-ray induced background in the xenon from each of the components contributing to it, as determined from the ZEPLIN--III data, have been performed.  
The weighted average tagging efficiency is predicted to be (27.0$\pm$0.6)\%, which, when supplemented by the 0.4\% accidental coincidence rate, rises to (27.4$\pm$0.6)\% and is consistent with the measured value.  
Since much of the $\gamma$-ray background in the xenon target arises in components beneath the xenon~\cite{backgrounds}, 
low energy shallow angle single scatters are preferentially tagged by the roof modules.  
Additionally, these modules are shorter than those in the barrel such that equivalent energy depositions result in larger photoelectron signals in the former.  
As a result, the efficiency for rejecting $\gamma$-ray background is dominated by the roof.

\begin{figure}[ht]
\includegraphics[width=8.6cm]{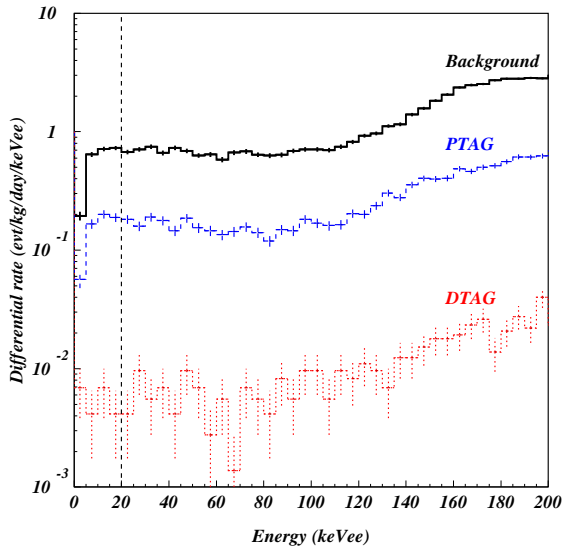}
\caption{\label{tagsDru}Differential background energy spectrum in the fiducial target of ZEPLIN--III in the second science run (solid line - labelled `Background').  
This rate is over an order of magnitude lower than the first science run as a result of the upgrades to the instrument.  
Overlaid are histograms of the differential rates of all events tagged as only prompt coincidences (dashed line - labelled `PTAG'), constituting approximately 28\% of the total rate, 
and events tagged as only delayed coincidences (dotted line - labelled `DTAG').  
The latter population is consistent with accidentally coincident delayed tags (as described in Section~\ref{neutrontagging}), and represent 0.7\% of the  electron recoil background.  
The vertical line represents the approximate upper boundary of the WIMP acceptance window.}
\end{figure}

The prompt tagging efficiency has been explored as a function of the energy of the coincident signal seen in ZEPLIN--III.  
Figure~\ref{tagsDru} shows the differential rate of background events in ZEPLIN--III for energies up to 200~keV$_{ee}$.  
The rate of electron recoils in the liquid xenon WIMP target (6.5~kg fiducial region) is 0.75$\pm$0.05 events/kg/day/keV at low energy, 
which represents a 20-fold improvement over the rate observed in the first run of the experiment~\cite{backgrounds}.  
The prompt tagging differential rate is approximately a constant fraction of the total.  
For synchronised signals depositing less than 20~keV$_{ee}$ in ZEPLIN--III (the approximate upper boundary of the WIMP acceptance region), the average fraction is (28.2$\pm$0.6)\%.

This efficiency represents a significant improvement over the veto system used in the ZEPLIN--I and ZEPLIN--II experiments~\cite{zep1,zep2si,zep2sd}.  
This is as a result of successful operation at the low threshold of only 2 photoelectrons (equating to $\sim$40 keV of energy deposition in the scintillator) 
and a narrow coincidence window mitigating the effects of background rate in a large tonne-scale external veto device that also doubles as shielding.  
In addition to background rejection, the significant $\gamma$-ray tagging efficiency further enhances the effectiveness of the veto detector as a diagnostic aid.  
In particular, it allows definition of an unbiased sample of background events for detector characterisation in a blind analysis and provides an independent estimate of the $\gamma$-ray background in the xenon target.  
This is especially important at low energies where particle discrimination is not perfect.  
Furthermore, as is discussed in Section~\ref{implications}, if a small population of WIMP candidates is found in ZEPLIN--III, the (lack of) prompt tagging can rule out a significant $\gamma$-ray component.  
Finally, a discrepancy between predicted and observed tagging efficiencies in the science exposure could indicate that electron recoil backgrounds have a significant $\beta^{-}$ contribution, 
such as that expected from $^{85}$Kr or surface contamination.

\section{\label{neutrontagging}Neutron tagging}

Single low-energy nuclear recoils from elastic neutron scattering in ZEPLIN–-III would be indistinguishable from WIMP interactions.  
A high efficiency neutron veto detector, however, mitigates against this irreducible background.  
Most neutrons generating a signal in the xenon target are effectively moderated to thermal energies in the polypropylene shielding and undergo radiative capture by the Gd.  
The detection of the capture $\gamma$-rays by the veto modules, placed outside of the Gd-loaded polypropylene, 
results predominantly in delayed pulses relative to the original scatter in the xenon.  
As mentioned previously, these veto events are labelled as `delayed' tags.  
Both the capture time distribution and the overall neutron tagging probability depend on the average Gd concentration and its spatial distribution in the hydrocarbon shield.  
The performance of the veto as an effective neutron detector relies more on those factors than on the scintillation yield, as discussed in Ref.~\cite{vetoPaper1}.

\subsection{\label{gdconcentration}Measurement of the mean capture time}

The veto detector was calibrated {\em in situ} using an Am-Be ($\alpha$,n) source inserted through the shieding to a position above the ZEPLIN--III instrument.  
This calibrates simultaneously the response of the WIMP target to nuclear recoils (as produced by WIMPs) and the veto detector efficiency for tagging internal background neutrons.  
The presence of Gd in the polypropylene increases the average number of $\gamma$-rays emitted per capture and shortens the mean capture time relative to that in the bare hydrocarbon shielding.  
This raises the detection efficiency and reduces the time window required to search for delayed coincidences with the target, respectively.  
It is important to measure accurately the time distribution for captures relative to neutron scattering in the xenon, 
since this validates the spatially-averaged concentration of Gd as well as the contribution of captures in other materials.

\begin{figure}[ht]
\includegraphics[width=8.6cm]{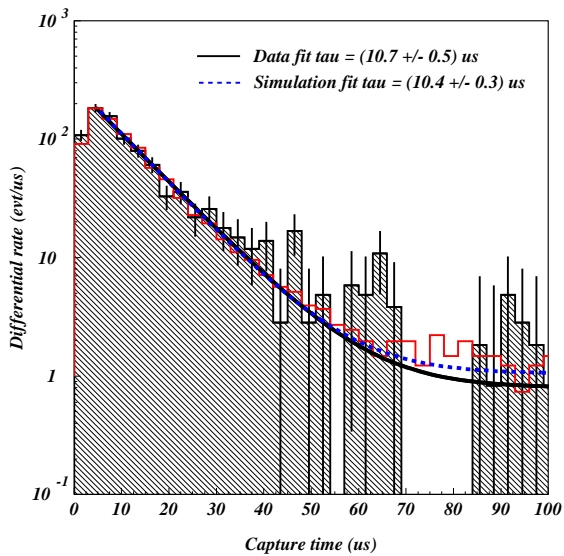}
\caption{\label{sim_comp_cg}Distributions of pulse times in the veto array relative to the {\em S1} signal from nuclear elastic scatters in ZEPLIN--III, 
for events that satisfy neutron selection described in the text (filled histogram).  
The fit to the data includes an exponential time delay distribution and a constant, representing accidental coincidences during calibration.  
The overall trend (solid line) has a characteristic decay time of (10.7$\pm$0.5)~$\mu$s. 
This is in excellent agreement with Monte Carlo simulations (open histogram), which yield a mean capture time of (10.4$\pm$0.3)~$\mu$s for a 0.42\% (w/w) Gd concentration~\cite{vetoPaper1}.}
\end{figure}

Nuclear recoils in the liquid xenon are identifiable by their typically lower {\em S2}/{\em S1} ratio relative to electron recoils.  
A population of single elastic scatters from neutrons is defined by selecting a nuclear recoil band within 2 standard deviations of the median.  
No additional cuts are applied except geometrical ones required to fiducialise the liquid xenon volume.  
Synchronised veto detector events are then searched for delayed signals that satisfy the selection criteria for neutron events and a neutron tagging efficiency is established, 
as described in Section~\ref{neutrontaggingefficiency}.  
The distribution of pulse times for these events in the veto detector relative to the time of the {\em S1} signal in the xenon is shown in Figure~\ref{sim_comp_cg}, 
along with Monte Carlo simulation results for comparison.  
The mean capture time is measured at (10.7$\pm$0.5)~$\mu$s, which corresponds to a Gd concentration of (0.42$\pm$0.03)\% (w/w)~\cite{vetoPaper1}.  
It has been determined that a range of 0.3-0.5\% will vary the neutron tagging efficiency by less than 1\%, 
and a measured content of 0.42\% satisifies this design specification.  

\subsection{\label{neutrontaggingefficiency}Neutron tagging efficiency}

The design of the veto has been driven by a requirement to maximise the rejection of neutron-induced background nuclear recoils in ZEPLIN--III 
that might otherwise be misidentified as WIMPs.  
Veto pulse selection criteria must maximise the neutron tagging efficiency whilst maintaining a low accidental coincidence rate in science data.  

The distribution of pulse times in the veto detector measured from the {\em S1} time in the xenon (Figure~\ref{sim_comp_cg}), implies that over 99\% of pulses arrive within a 70~$\mu$s window.  
The differential rate of events drops by over two orders of magnitude up to this point and, for this neutron calibration exposure, very few de-excitations occur beyond this timescale (sampling limited by statistics).  
The characteristic timescale obtained is consistent with a Gd concentration of 0.3-0.5\% by weight within the polypropylene shielding~\cite{vetoPaper1}.  
Beyond 70~$\mu$s the tagging efficiency increases slowly, but  only due to (expected) accidental coincidences.  
At short capture times, a small fraction of neutrons will undergo capture even within the small prompt coincidence window (described in Section~\ref{gammatagging}).  
While these neutron events will still be rejected by the veto, they will be excluded from the delayed tag efficiency.  
This contribution to the overall neutron tagging efficiency of the veto must also be considered.

The choice of threshold for the delayed tag, in terms of both number of photoelectrons and multiplicity, has been set so as to limit the accidental coincidence tags in background data to a maximum of 1\%.  
A rate in excess of this poses an unacceptable loss of effective exposure given the neutron-induced nuclear recoil expectation in the WIMP acceptance region of less than 1 event per year following the upgrades to the internal PMT array.  
The desired threshold is achieved by relaxing the neutron selection criteria in the veto detector until 1\% of electron recoil events in ZEPLIN--III from background data are accidentally labelled delayed tags.  
This is achieved with a 10 photoelectron minimum threshold in the veto detector.  
No minimum multiplicity condition is applied.  
As with prompt tags with pulses distributed across the veto detector, all pulses contributing to a delayed tag must be coincident with all others within a $\pm$0.2~$\mu$s window.  

\begin{figure}[ht]
\includegraphics[width=8.6cm]{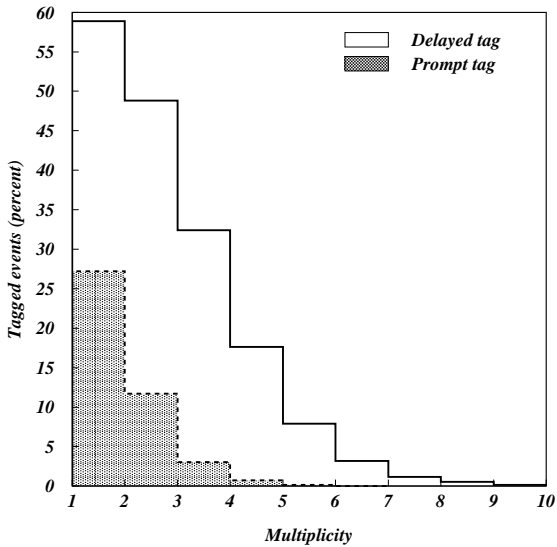}
\caption{\label{nvg}Absolute prompt and delayed tagging efficiencies as a function of veto module multiplicity.  
Here the photoelectron threshold is held constant at 2 photoelectrons for prompt tags and 10 photoelectrons for delayed tags, and only the multiplicity requirement varied.  
Increasing the multiplicity requirement from 1 to 2 for the delayed tag reduces the efficiency by $\sim$10\% for a fixed 10 photoelectron threshold.  
In contrast to prompt tags, if a higher multiplicity were adopted, such a loss in efficiency could be recouped by using a lower threshold with no increase to 
delayed accidental coincidences.}
\end{figure}

Figure~\ref{nvg} shows the absolute delayed tagging efficiency as a function of minimum required veto detector multiplicity, when the threshold is held constant.  
Increasing the multiplicity requirement from 1 to 2 lowers the efficiency by approximately 10\%.  
The 10 photoelectron threshold with no multiplicity is found to yield an efficiency and accidental coincidence rate equivalent to a multiplicity of 2 modules recording a total of at least 8 photoelectrons.  
The rate of accidental coincidences for delayed tags remains low despite a lower photoelectron threshold as a result of the stronger dependence on multiplicity for background $\gamma$-rays that make up the accidental rate.  
Combining the two selection criteria, {\em i.e.}, higher photoelectron threshold with no multiplicity and lower photoelectron threshold with a multiplicity of 2, 
does not result in a gain in efficiency since they each tag correlated datasets.  
For the same reason, however, the accidental coincidence rate is increased.  
As such, the delay tag selection criteria is only that a combined minimum of 10 photoelectrons equivalent signal be detected within the delay window, 
with no minimum multiplicity condition, although any pulses distributed across multiple modules must be in coincidence with all others.

Although the delay tag criteria have been set for an accidental rate of 1\%, it is reduced in background data by 0.28\% - the prompt tagging efficiency.  
This is because the population to which the delayed tag applies has already been decreased in exposure by the prompt tags, which are assessed first.
Consequently, the accidental coincidence rate for the delayed tag in the background data is only 0.7\%.  
The differential rate of these events in background data is shown in Figure~\ref{tagsDru}.  
The combined prompt and delayed tagging accidental coincidence rates for ZEPLIN--III background, 0.4\% and 0.7\%, respectively, is 1.1\%.  
The accidental tagging rate for WIMPs, however, remains 1.4\% (0.4\% from the prompt tag and 1\% from the delayed tag).

\begin{figure}[ht]
\includegraphics[width=8.6cm]{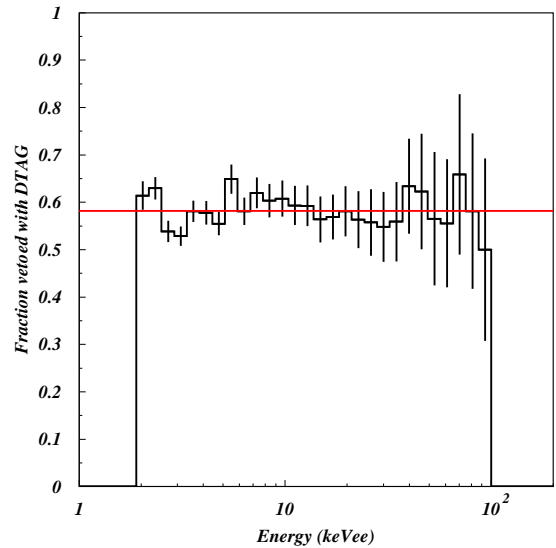}
\caption{\label{drufrac_100825_2sigma}The delayed neutron tagging efficiency as a function of nuclear recoil energy deposition in the liquid xenon target.  
The average tagging efficiency is (58.8$\pm$0.5)\% for less than 20~keV$_{ee}$ energy deposition single scatters in the WIMP acceptance window in ZEPLIN--III.  
This efficiency remains constant since the probability for detection of $\gamma$-rays following radiative capture of the neutron and the de-excitation of $^{158}$Gd is independent of 
the neutron energy following the scatter in the xenon.}
\end{figure}

Figure~\ref{drufrac_100825_2sigma} shows the fraction of single scatter nuclear recoils within ZEPLIN--III from an AmBe neutron calibration dataset 
that are accompanied by a delayed tag in the veto detector, as a function of energy deposition in the xenon target.  
The delayed tag threshold is as described earlier, namely a 10 photoelectron threshold with no minimum multiplicity requirement.  
The delayed tagging efficiency accurately reproduces the spectral shape of the nuclear recoils, with no significant deviations, at a constant fraction close to 59\%.  
The mean delayed tagging efficiency for these single scatter nuclear recoils depositing less than 20~keV$_{ee}$ in the WIMP acceptance window is (58.8$\pm$0.5)\%.  
The efficiency remains constant with ZEPLIN--III energy since the probability of detection of the delayed $\gamma$-rays from the de-excitation of the $^{158}$Gd nucleus is independent of the original neutron energy.  
This is because the angular distribution of the interacting neutrons in the xenon is destroyed through proton recoils in the polypropylene and thermalisation prior to capture.

For a definition of the full neutron tagging efficiency, the delayed tagging fraction must be supplemented by the fraction of neutron coincidences falling in the prompt window.  
As stated previously, these events would be vetoed by the prompt tag.  
However, it is not possible to measure this fraction using the neutron calibration data with exposure to an AmBe neutron source.  
This is due to the emission of high energy $\gamma$-rays from the AmBe source in coincidence with the neutron emitted in the Be($\alpha$,n) reaction.  
Such $\gamma$-rays are of course detected in the prompt acceptance window and provide accidental tags for neutron scatters in the xenon target.  
Moreover, due to their high energies (mostly 4.44~MeV), they are not representative of $\gamma$-ray background and are indistinguishable in size and multiplicity from the delayed $\gamma$-rays from 
de-excitation of the $^{158}$Gd nucleus following radiative capture.  
Consequently, the population of neutron events in the prompt acceptance window cannot be measured directly, 
but it may instead be calculated by extrapolating the delayed time distribution for neutron events into the positive half of the prompt window.  
Given the characteristic time of the delayed tag time distribution, an additional (1.7$\pm$0.1)\% in neutron tagging efficiency is to be expected from the prompt window.  
Monte Carlo simulations of the neutron calibration exposure predict (1.5$\pm$0.1)\% for the equivalent window, 
and a total neutron tagging efficiency (60.7$\pm$0.1)\% for a 10 photoelectron threshold~\cite{vetoPaper1}.  
The measured combined (prompt+delayed window) neutron tagging efficiency is (60.5$\pm$0.5)\% for the same threshold, in excellent agreement with simulations.  
With a 60\% neutron tagging efficiency, the number of expected nuclear recoil events per year from neutron background in the 
WIMP acceptance region of ZEPLIN--III ({\em i.e.}, 6.5~kg fiducial region, $\sim$5-50~keV) indistinguishable from WIMP signal, is reduced to $\simeq$0.2.

\section{\label{implications}Implications for signal limits}

\begin{figure}[ht]
\includegraphics[width=8.6cm]{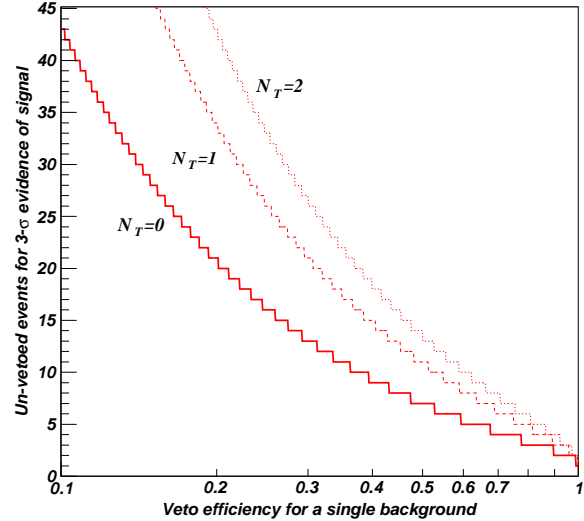}
\caption{\label{veto_power}The effect of veto efficiency on discovery power for a generic rare
event search with a single background and no additional discrimination.
$N_{\mathrm{T}}$ is the number of tagged events observed, and confidence
intervals are calculated as in Ref.~\cite{PLR}.}
\end{figure}

If one considers vetoed events as a measurement of the rate of un-vetoed background events with Poisson uncertainty, 
that information can be incorporated into limits on the signal rate.  
The efficiency, $\eta$, for vetoing a background event gives the relative exposure of vetoed to un-vetoed background samples: $\frac{\eta}{1-\eta}$.  
A confidence interval for signal can then be set using the profile likelihood ratio (PLR)~\cite{PLR} as implemented in the ROOT~\cite{ROOT} class TRolke.  
For a single-background case, Figure~\ref{veto_power} shows the number of untagged events that would constitute 3$\sigma$ evidence for signal, 
as a function of $\eta$.  
With $\eta$=0.28, it is seen that 15 un-vetoed events (and no vetoed events) in a search region would be sufficient;
this number falls for higher veto efficiencies, eventually reaching 1.  
In ZEPLIN--III and similar experiments, nuclear- and electron-recoil background rates must be summed and have different veto efficiencies; 
however, simulations and measurements away from the search region constrain the neutron background more tightly than an absence of 
delayed tagged events in the search box alone.  
Additionally, the signal and the electron-recoil background -- and, to a lesser extent, the nuclear-recoil background -- 
are differently distributed in more parameters than just veto tagging efficiency, 
for example pulse-shape, energy, multiplicity and {\em S2}/{\em S1}.  
By using these additional discriminants, one could reject the background-only hypothesis with even fewer un-vetoed, 
signal-like events than the simple counting case of Figure~\ref{veto_power}.

\section{Summary}
\noindent 

The veto detector has been operational for over 6 months with stable performance.  
The duty cycle of the veto detector is 100\% to that of the ZEPLIN--III instrument, and data streams from both detectors have been synchronised successfully with near unity efficiency.  
The interaction rates in the veto detector are in excellent agreement with expected background based on dedicated measurements of all detector components, 
and this validates the impact on both electron and nuclear recoil event rates in ZEPLIN--III.  
Daily SPE and weekly LED calibrations have been supplemented with neutron exposures to determine energy scales and tagging efficiencies for both $\gamma$-ray induced prompt, 
and neutron-induced delayed coincidences.  

Selection criteria for both prompt and delayed tagging have been defined such that the combined accidental coincidence fraction is approximately 1\% in ZEPLIN--III.  
The veto detector rejects over 28\% of coincident electron recoil background in the xenon target, improving rejection of leakage $\gamma$-rays, providing diagnostic information for any spurious event populations, 
and supporting background expectation analysis.  
The neutron tagging efficiency of the veto detector is such that 60.5\% of the expected nuclear recoil background in the second science run of the experiment can be identified.  
After veto tagging, the neutron event expectation in a 1 year long dataset with typical signal acceptance is some 10 times lower than the pre-upgrade levels of the first science run~\cite{z3fsr}, 
and ZEPLIN--III could achieve a sensitivity of $\sim$1$\times$10$^{−8}$~pb$\cdot$year to the scalar WIMP-nucleon elastic cross-section~\cite{backgrounds}.  
The veto detector continues to operate in the ongoing second science run of the ZEPLIN--III experiment.

\begin{acknowledgments}

The UK groups acknowledge the support of the Science \& Technology Facilities Council (STFC) for the ZEPLIN--III project and for maintenance and operation of the underground Palmer 
laboratory which is hosted by Cleveland Potash Ltd (CPL) at Boulby Mine, near Whitby on the North-East coast of England.  
The project would not be possible without the co-operation of the management and staff of CPL. 
We also acknowledge support from a Joint International Project award, held at ITEP and Imperial College, from the Russian Foundation of Basic Research (08-02-91851 KO\_a) and the Royal Society.
LIP--Coimbra acknowledges financial support from Funda\c c\~ao para a Ci\^encia e Tecnologia (FCT) through the project-grants CERN/FP/109320/2009 and CERN/FP/116374/2010, 
as well as the postdoctoral grants SFRH/BPD/27054/2006, SFRH/BPD/47320/2008 and SFRH/BPD/63096/2009.  
This work was supported in part by SC Rosatom, contract $\#$H.4e.45.90.10.1053 from 03.02.2010.  
The University of Edinburgh is a charitable body, registered in Scotland, with the registration number SC005336.

\end{acknowledgments}

\end{document}